# Structural optimization of lattice-matched $Sc_{0.14}Al_{0.86}N$/GaN superlattices for photonic applications

Rajendra Kumar[1,2], Govardan Gopakumar[1,2], Zain Ul Abdin[1,2], Michael J. Manfra[1,2,3,4], and Oana Malis[1,2]*

[1]Dept. of Physics and Astronomy, Purdue University, West Lafayette, IN USA 47907

[2]Birck Nanotechnology Center, West Lafayette, IN USA 47907

[3]School of Materials Engineering, Purdue University, West Lafayette, IN USA 47907

[4]Elmore Family School of Electrical and Computer Engineering, Purdue University, West Lafayette, IN USA 47907

**Abstract**

$Sc_xAl_{1-x}N$ is an emerging III-nitride material known for its high piezoelectric coefficient and ferroelectric properties. Integration of wide-bandgap $Sc_xAl_{1-x}N$ with GaN is particularly attractive for quantum photonic devices. Achieving low defect complex multilayers incorporating $Sc_xAl_{1-x}N$, though, requires precise lattice-matching and carefully optimized growth parameters. This study systematically investigates the molecular-beam epitaxy of short-period $Sc_xAl_{1-x}N$/GaN superlattices with total thicknesses of up to 600 nm on GaN templates. X-ray diffraction reciprocal space mapping confirmed lattice-matching at $x = 0.14 \pm 0.01$ Sc composition regardless of the thickness of GaN interlayers, as evidenced by symmetric superlattice satellites aligned in-plane with the underlying substrate peak. Superlattices with Sc compositions deviating from this lattice-matching condition exhibited strain-induced defects ranging from crack formation to partial relaxation. Scanning transmission electron microscopy (STEM) investigation of the $Sc_xAl_{1-x}N$/GaN interfaces identified temperature-dependent intermixing as a major factor in setting the nitride composition variation and implicitly band structure profile along the growth direction. Energy-dispersive X-ray spectroscopy also revealed that Sc incorporation exhibits delays relative to Al at both onset and termination. Optimal growth conditions were observed at approximately 600°C and 550°C for superlattices with thick GaN layers (6 nm), and ultra-thin GaN layers (≤ 2 nm), respectively.

---

* Author to whom correspondence should be addressed. Electronic mail: omalis@purdue.edu

## I. INTRODUCTION

Scandium aluminum nitride ($Sc_xAl_{1-x}N$) has emerged as a highly promising nitride semiconductor, garnering considerable research attention over the past decade. This is in part due to significantly higher piezoelectric coefficients of $Sc_xAl_{1-x}N$ alloys compared to AlN.[1,2] Due to the high electromechanical coupling coefficient of $Sc_xAl_{1-x}N$, its film-based bulk acoustic wave (BAW) resonator filters are particularly promising for high-frequency signal processing and have already been commercialized for mobile communications.[3] Notably, $Sc_xAl_{1-x}N$ BAW resonators are now commonly integrated into a wide range of smartphones, underscoring the material's maturity and industrial relevance. Furthermore, the ferroelectric properties of $Sc_xAl_{1-x}N$ grown by molecular beam epitaxy have enabled novel device concepts such as ferroelectric memories and high-electron mobility transistors.[4–9]

Beyond piezoelectric and ferroelectric applications, the integration of $Sc_xAl_{1-x}N$ into III-nitride heterostructures has opened exciting avenues for advanced photonic devices, especially those operating in the ultraviolet and infrared spectral regions. $Sc_xAl_{1-x}N$ alloys, with tunable wide bandgap energies and lattice constants, are particularly attractive for optoelectronic devices, such as lasers and photodetectors,[10,11] due to their ability to achieve lattice-matching with GaN. However, the successful realization of high-performance devices relies on the meticulous epitaxial growth of complex layer sequences with accurately controlled thicknesses and composition profiles. In particular, precise lattice-matching of $Sc_xAl_{1-x}N$ with GaN is critical for low-defect thick nitride stacks with excellent structural integrity. Our earlier work[12] established x = 0.14±0.01 as the optimal composition for strain-free $Sc_xAl_{1-x}N$ on GaN using high-resolution X-ray diffraction and Rutherford backscattering spectroscopy (RBS). Building on this foundation, this work systematically investigates the impact of varying Sc compositions, growth temperatures and layer thicknesses on the structure of $Sc_xAl_{1-x}N$/GaN short-period superlattices (SL) grown by plasma-assisted molecular beam epitaxy (MBE).

The growth of high-quality SLs presents multifaceted challenges, beginning with stringent strain management requirements; small deviations from the lattice-matched Sc composition induce cumulative strain, severely limiting the critical thickness before relaxation via cracking or plastic deformation. Compounding this is a fundamental thermal trade-off: while high-quality (Al)GaN typically requires MBE growth temperatures above 700°C, ScAlN layers begin to exhibit structural degradation above 600°C.[13,14] This necessitates precise optimization of the substrate temperature to navigate these competing requirements. Moreover, atomic-scale interface control is essential, as interdiffusion at the ScAlN/GaN interfaces leads to quantum well broadening and compromises the performance of devices that rely on abrupt interfaces.

In this study, we have employed high-resolution x-ray diffraction and scanning transmission electron microscopy techniques to evaluate the structural properties of $Sc_xAl_{1-x}N$/GaN SLs and systematically tackle the challenges mentioned above. We demonstrate that SLs with $x = 0.14$ are extended defect-free and coherently strained with the underlying GaN substrate. We reveal how GaN layer thickness modulates strain relaxation dynamics in compressively strained structures and quantify interfacial intermixing at the monolayer scale. Particular attention is given to the effect of growth temperature on the structure of interfaces. These results identify key growth parameters for achieving high-quality $Sc_{0.14}Al_{0.86}N$/GaN SLs suitable for ultraviolet and infrared photonic applications.

## II.    EXPERIMENTAL METHODS

50-period superlattices consisting of GaN quantum wells (QW) and ScAlN barriers with varying Sc compositions and well thicknesses were grown using plasma-assisted molecular beam epitaxy on commercially supplied Fe-doped semi-insulating c-plane GaN templates (GaN thickness of 2 μm, dislocation density of $8\times10^8$ cm$^{-2}$) on sapphire (0001) substrates. The substrates were backside coated with a 1 μm layer of tungsten silicide to improve the thermal coupling between substrate and the MBE heater. The substrate cleaning procedure was described elsewhere.[10] The substrates were outgassed overnight for more than 12 hours at 550°C in an ultra-high vacuum chamber connected to the MBE system. To minimize impurity-related issues in the SLs, high-purity metal sources were used for Al, Ga (purity 99.99999%) and Sc (purity 99.98% trace rare-earth metals). Active nitrogen was supplied by a Veeco Unibulb radio-frequency (RF) plasma source, operating at a forward power of 250 W. The N$_2$ flow rate was maintained at 0.25 sccm which corresponds to a GaN growth rate of approximately 3.6 nm/min. Before the growth of the SL, a 150 nm GaN buffer layer was grown at a substrate temperature of 740°C under Ga-rich conditions, resulting in a smooth surface with a root-mean-square (rms) roughness of 0.4 nm over a 5 μm × 5 μm area. For the superlattice growth, the substrate temperature ($T_{sub}$) measured by pyrometer was varied between 500°C and 625°C. The metal-to-nitrogen ratios for the ScAlN and GaN layers in the superlattice were approximately 0.7 and 1.2, respectively. The growth rate of ScAlN layers in the SL is 2.6 nm/min. The structures contain silicon doping for optical studies (not included here) provided by two δ-doping sheets placed 1 nm away from each interface within the barriers. The doping density was controlled by the duration of the δ-doping process. A nominal 10-second δ-doping sheet corresponded to either silicon density of $4.0 \times 10^{12}$ cm$^{-2}$ for $T_{Si}$=1375°C or $8.0 \times 10^{12}$ cm$^{-2}$ for $T_{Si}$=1402°C.

The structural characterization of the samples was conducted using atomic force microscopy (AFM), X-ray diffraction (XRD), and scanning transmission electron microscopy. High-angle annular dark-field STEM (HAADF-STEM) images were used to probe the local atomic structure of individual layers. HAADF-STEM samples were prepared using a Thermo Scientific Helios G4 UX dual focus ion beam (FIB) system, followed by cleaning with a Fischione Ar plasma cleaner to remove contamination from the FIB process or air exposure. Imaging and analysis were performed with a Thermo Scientific Themis Z double aberration-corrected S/TEM, operating at an acceleration voltage of 300 kV and a screen current of 0.12 nA. Energy-dispersive X-ray spectroscopy (EDX) was performed using the Super-X quad silicon drift detector system installed in the Themis Z. A screen current of ~0.5 nA was used to enhance the intensity of characteristic X-ray emission, and elemental maps were acquired with a dwell time of 5 μs per pixel. The acquired EDX maps were denoised using principal component analysis (PCA), applied to the entire image. The number of components retained was selected to minimize apparent Al and Sc signals within the GaN buffer region, where their presence is not expected.

Geometric phase analysis (GPA) of HAADF-STEM images was employed to quantify the lattice strain between GaN and $Sc_xAl_{1-x}N$ layers. GPA enables quantitative mapping of strain and lattice distortions in HAADF-STEM images by analyzing the Fourier components of a target region relative to a reference region.[15] The analysis was conducted using an open-access plugin[16] for Digital Micrograph™ (Gatan Inc.).

High-resolution X-ray diffraction (HRXRD) measurements of the (0002) and (10$\bar{1}$5) reflections were carried out using a Panalytical Empyrean High-Resolution diffractometer. The Empyrean diffractometer is equipped with a PIXcel$^{3D}$ detector, which can operate as both point (0D) and one-dimensional (1D) detector. Reciprocal space maps (RSMs) were obtained using the 1D configuration for rapid mapping, while symmetric ω-2θ scans of the (0002) reflection were performed in the 0D configuration. RSMs of the (10$\bar{1}$5) reflection provide information about the strain state of the SL.

Since it is not feasible to perform RBS or secondary ion mass spectrometry (SIMS) on very thin layers, we used HRXRD simulations to estimate average Sc composition and layer thicknesses in the SLs. Noteworthy, STEM EDX also exhibits relatively high error margins (~2-4%) in Sc composition estimation, limiting its effectiveness in distinguishing small compositional variations between SLs. Therefore, the well and barrier thicknesses and Sc compositions listed in the text were calculated from fitting symmetric ω-2θ scans of the (0002) HRXRD reflection with the X'pert Epitaxy software assuming the ScAlN lattice constants proposed in our previous studies.[12,13] We note that the fitted Sc-compositions are consistent with expectations from beam-flux measurements calibrated with HRXRD measurements of thin ScAlN films calibrated using RBS measurements.[12]

### III. RESULTS AND DISCUSSION

Even though the exact lattice-matching composition for $Sc_xAl_{1-x}N$ on GaN has been reported in the literature at values ranging from 9% to 18%,[12,13,17–23] our previous systematic analysis with HRXRD and RBS of 100-nm thick $Sc_xAl_{1-x}N$ films on GaN presented compelling experimental evidence for lattice-matching conditions at x = 0.14±0.01.[12] To further test this hypothesis, we compare the structure of 50-period 6-nm-$Sc_xAl_{1-x}N$/6-nm-GaN SL samples with Sc composition x = 0.12 - 0.18 using XRD RSM, as shown in Figure 1. The RSM of the $Sc_{0.14}Al_{0.86}N$/GaN sample exhibits narrow and symmetric SL satellites aligned in the $Q_x$ direction with the underlying GaN template, confirming the in-plane lattice constant is maintained at the value of GaN throughout the 600-nm thick superlattice. We note that the tilted peak shape in reciprocal space is caused by experimental resolution as best illustrated by the GaN substrate reflection. In contrast, the XRD RSM of the $Sc_{0.12}Al_{0.88}N$/GaN sample shows SL satellites that are asymmetric towards higher $Q_x$ indicating that x = 0.12 is below the lattice-matching composition. This is corroborated by observation of cracks in optical microscopy (see Figure S1 in Supplementary Material), and is in agreement with previous reports showing that layers under tensile strain tend to relax through crack formation.[24] Surprisingly, the RSM of the $Sc_{0.18}Al_{0.82}N$/GaN sample also displays narrow satellites indicating the SL is coherently strained to GaN. Based on our previous work on 100-nm $Sc_{0.18}Al_{0.82}N$ films,[12] we expected this sample to show plastic relaxation through broadening and shifting of the satellite peaks. However, the critical thickness of compressively-strained multilayers that can be grown before onset of plastic relaxation is significantly affected by the presence and quality of unstrained layers (GaN).[24] Therefore, we attribute the unexpected stability of this sample to the presence of low-defect 6 nm GaN QWs that modulate strain throughout the superlattice.

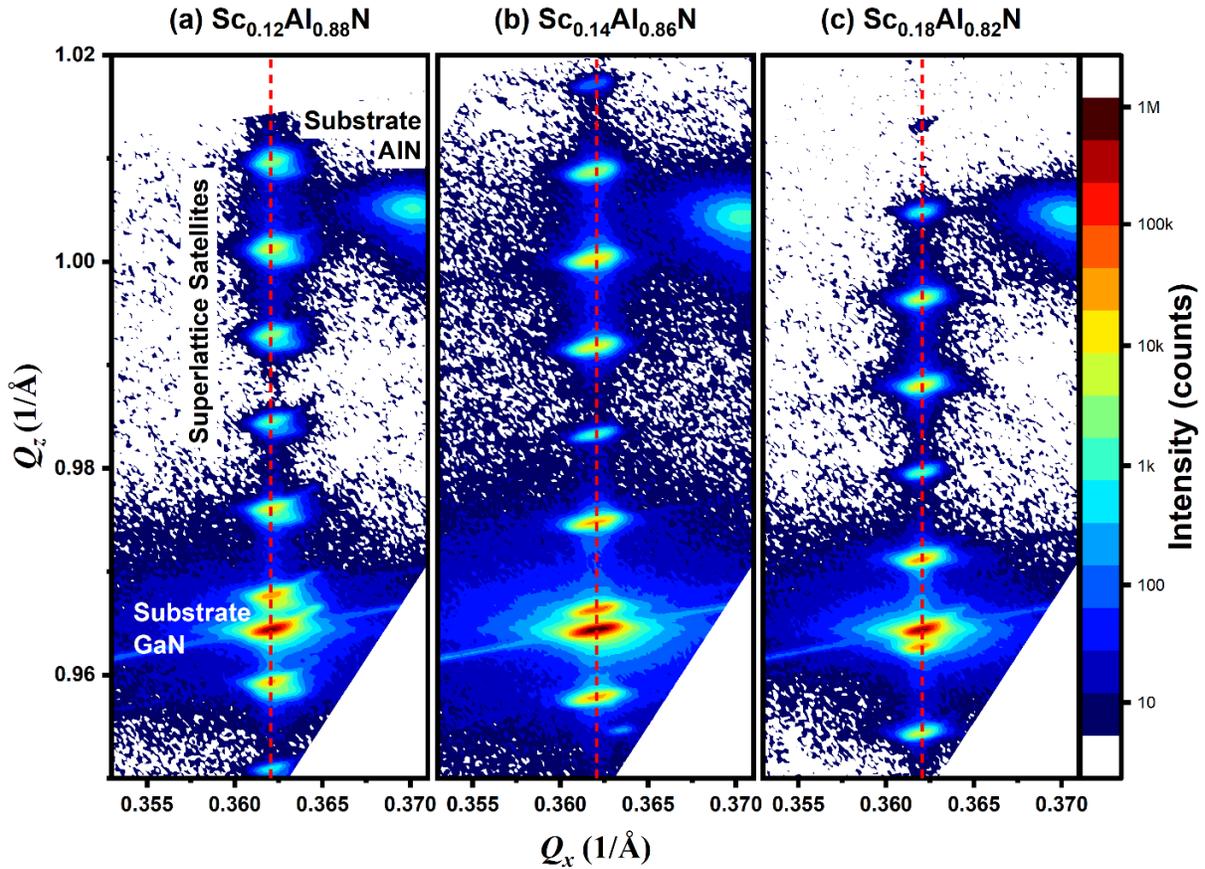

**Figure 1.** XRD RSM plots of the $(10\bar{1}5)$ reflection for 50-period 6-nm-$Sc_xAl_{1-x}N$/6-nm-GaN SLs with Sc composition of (a) 12%, (b) 14%, and (c) 18%. The samples were grown at 600°C. Vertical dashed red lines indicate the position corresponding to the in-plane GaN lattice constant. The most intense peak at $Q_z = 0.964$ corresponds to GaN template. Additional peak at $Q_x = 0.37$ and $Q_z = 1.00$ corresponds to an AlN interlayer between GaN template and sapphire substrate.

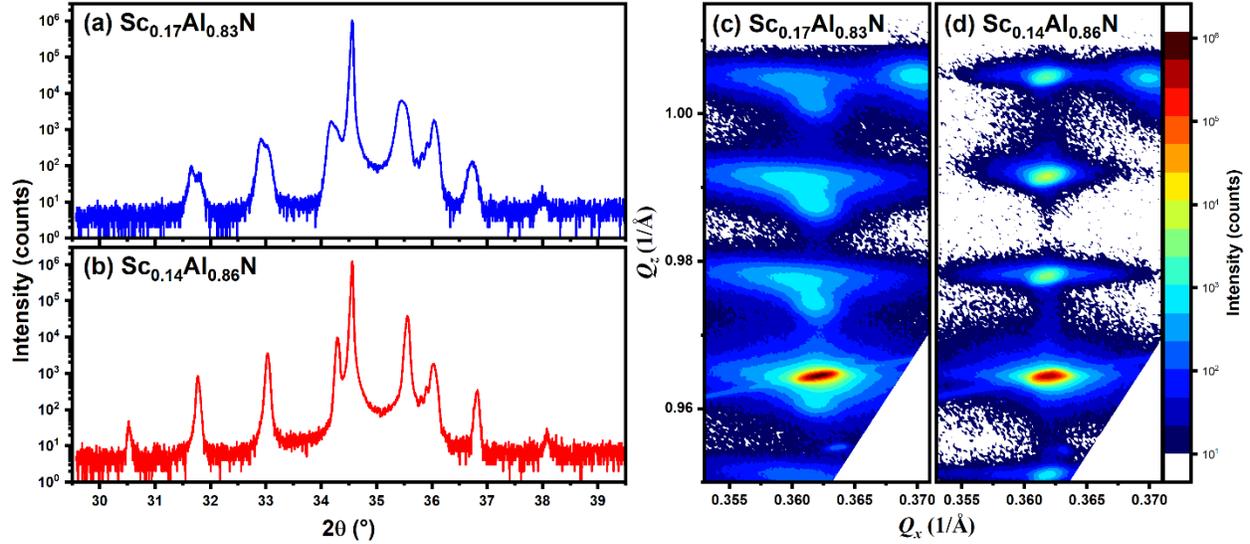

**Figure 2.** (a) HRXRD ω-2θ scan and (c) XRD RSM of the $(10\bar{1}5)$ reflection for a 6-nm-$Sc_{0.17}Al_{0.83}N$/2-nm-GaN SL. (b) HRXRD ω-2θ scan and (d) XRD RSM of the $(10\bar{1}5)$ reflection for a 6-nm-$Sc_{0.14}Al_{0.86}N$/2-nm-GaN SL. In XRD RSM, the most intense peak at $Q_z = 0.964$ corresponds to GaN substrate. The first sample was grown at $T_{sub} = 520°C$ while the second sample was grown at 535°C.

High-quality low-mismatch multilayer structures containing unstrained layers have been known to exhibit much higher relaxation critical thickness than expected from simple strain calculations.[24] To explore the impact of unstrained layer thickness on compressive strain relaxation, we have grown $Sc_xAl_{1-x}N$/GaN SLs with different GaN thicknesses while maintaining the $Sc_xAl_{1-x}N$ barrier at 6 nm. The symmetric ω-2θ scans and XRD RSMs of two samples with 2 nm GaN wells are shown in Figure 2. Reducing the GaN thickness to 2 nm does not affect alignment of the $Sc_{0.14}Al_{0.86}N$/GaN SL satellites in the $Q_x$ direction, consistent with the assumption of lattice-matching between well and barrier materials. For the $Sc_{0.17}Al_{0.83}N$/GaN SL, though, two sets of satellites are observed in Figure 2(c): one corresponding to a coherently strained region and another to a relaxed region. This behavior is also reflected in the ω-2θ scans, which show broad reflections and/or two sets of peaks for $Sc_{0.17}Al_{0.83}N$/GaN SL (Figure 2(a)) as opposed to a single set for $Sc_{0.14}Al_{0.86}N$/GaN (Figure 2(b)). The thinner GaN layers in the $Sc_{0.17}Al_{0.83}N$/GaN SL lead to a reduced critical thickness, resulting in a higher degree of plastic relaxation compared to the SL with 6 nm GaN layers (Figure 1(c)). The gradual evolution of the degree of relaxation for $Sc_{0.18}Al_{0.82}N$/GaN SLs with decreasing GaN thicknesses is further documented in Figure S2 in Supplementary Material file. Table S1 summarizes the average lattice-mismatch extracted from

RSMs in Figure S2. It is important to emphasize that any degree of relaxation undermines the performance of SLs in addition to compromising mechanical stability and therefore limits their usage in practical photonic devices. For instance, infrared intersubband absorption measurements (see Figure S3 in Supplementary Material) show that the $Sc_{0.14}Al_{0.86}N$/GaN SL exhibits four times higher total absorption compared to a $Sc_{0.18}Al_{0.82}N$/GaN SL, underscoring the functional advantages of lattice-matched SLs.

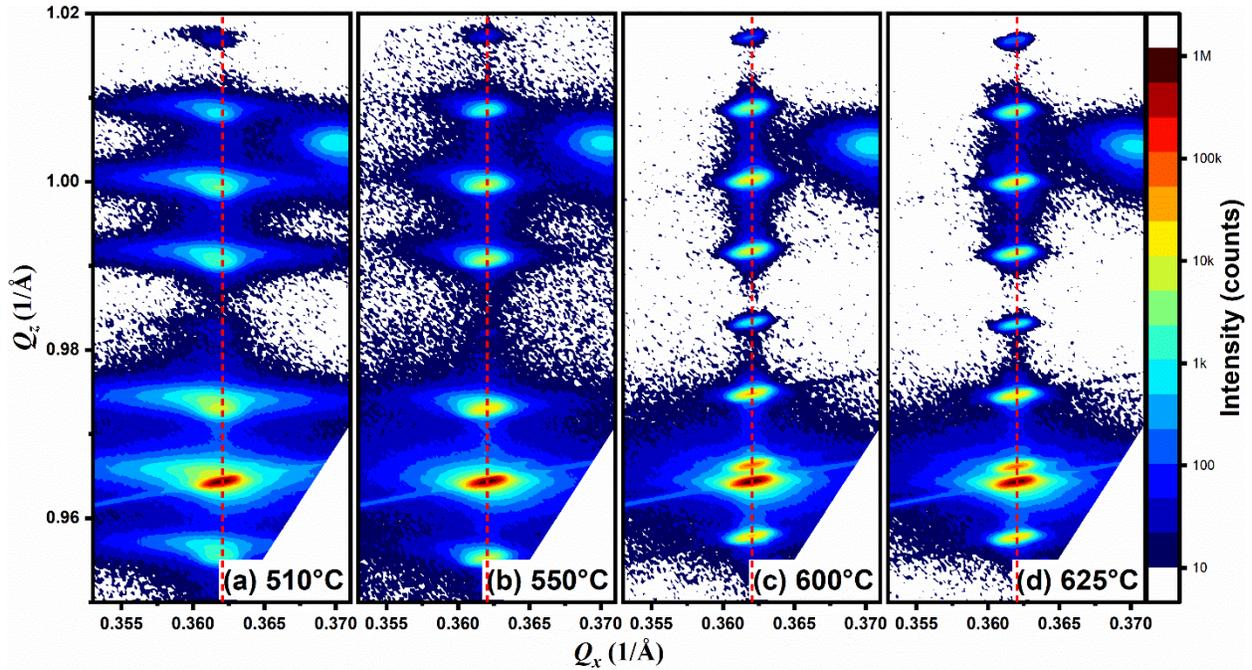

**Figure 3.** XRD RSM plots of the $(10\bar{1}5)$ reflection for 6-nm-$Sc_{0.14}Al_{0.86}N$/6-nm-GaN SLs grown at a substrate temperature of (a) 510°C, (b) 550°C, (c) 600°C and (d) 625°C. The most intense peak at $Q_z = 0.964$ corresponds to GaN substrate.

Substrate temperature plays a critical role in determining the structure of SLs by influencing the mobility of impinging metal atoms on the growth surface. While high-quality GaN films are typically grown at substrate temperatures above 700°C, $Sc_xAl_{1-x}N$ films tend to develop additional defects at these temperatures and are usually grown below 600°C.[13] To investigate the effect of substrate temperature on SL structure, we compared four lattice-matched 6-nm-$Sc_{0.14}Al_{0.86}N$/6-nm-GaN samples grown at different substrate temperatures, keeping all other growth parameters the same. The XRD RSM results, shown in Figure 3, confirm that all SLs are coherent in-plane with the GaN substrate. The sample grown at 510°C exhibits significant peak broadening in both $Q_x$ and $Q_z$ direction indicating relatively small coherent domain sizes. As the substrate temperature increased from 510°C to 600°C the SL satellites became narrower, with negligible differences between samples grown at 600°C and 625°C.

Line profiles of the SL peaks along $Q_x$ and $Q_z$ directions were extracted and fitted with Gaussian peaks, and the full-width-at-half-maximum (FWHM) values are summarized in Table S2 in the Supplementary Material file. The FWHM values decrease with increasing substrate temperature, indicating an improvement in the crystallinity of the SLs. A similar trend was observed in the linewidth of the SL fringes in HRXRD $\omega$-2$\theta$ scans (see Figure S4 and table S3 in Supplementary Material). Additionally, atomic force microscopy (AFM) measurements (see Figure S5 in Supplementary Material) revealed clear atomic step formation for SL samples grown at 600°C and 625°C substrate temperature, indicating excellent surface morphology of these SLs. In contrast, no atomic steps were observed in the AFM image of SL sample grown at 510°C.

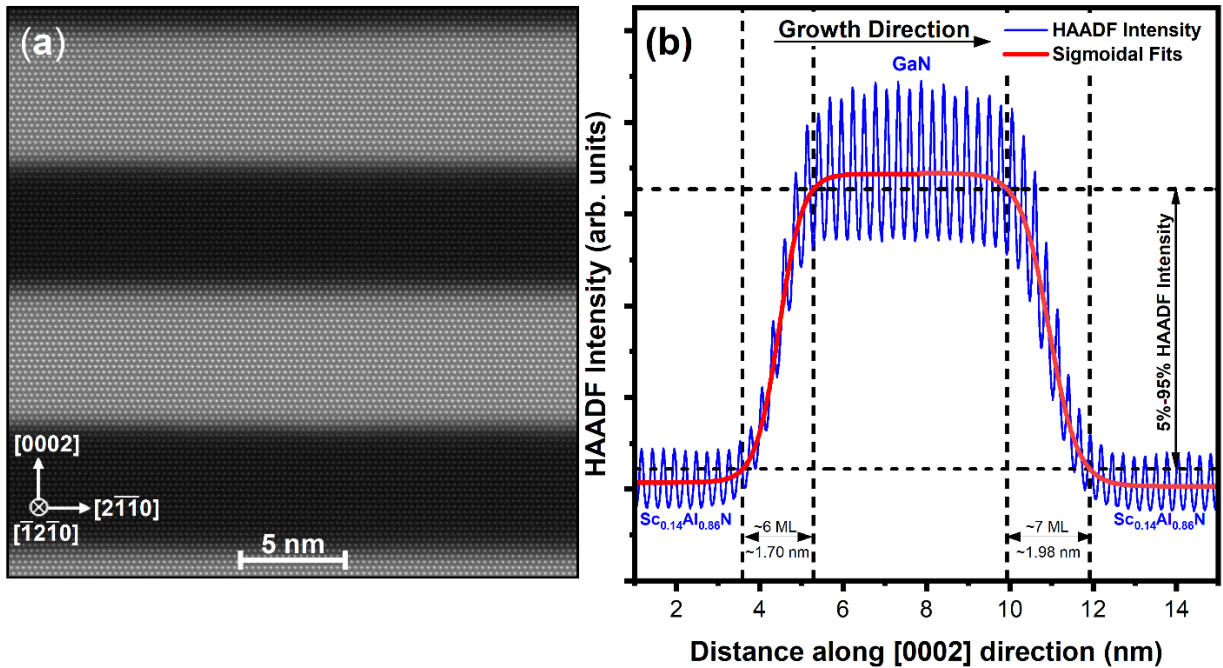

**Figure 4.** (a) HAADF-STEM image of 6-nm-$Sc_{0.14}Al_{0.86}N$/6-nm-GaN SL grown at 600°C substrate temperature. (b) HAADF-STEM intensity line profile along the growth direction across the bottom GaN QW averaged over 10 nm in the [$2\bar{1}\bar{1}0$] direction.

In quantum devices, interfaces control the band edge profile and ultimately play a major role in electronic and optical properties. To study the detailed atomic structure of the SL interfaces, we performed HAADF-STEM measurements complemented with STEM EDX maps (Figs. 4, 5, S6).

No extended defects were visible in low magnification STEM (Figure S6 in Supplementary Material). However, short-period superlattices are particularly susceptible to interfacial phenomena such as intermixing and roughening. To quantify these effects, we define the interface width as the distance over which the STEM intensity profile changes along the growth direction from 5% to 95% of the total signal contrast between the $Sc_{0.14}Al_{0.86}N$ and GaN layers. Figure 4 shows the HAADF-STEM image of a 6-nm-$Sc_{0.14}Al_{0.86}$N/6-nm-GaN SL grown at 600°C and the corresponding intensity profile of a QW extracted from the image by averaging over 10 nm in the $[2\bar{1}\bar{1}0]$ direction. To determine the interface widths, the intensity profile was fitted with sigmoidal functions. The bottom interface has a width of approximately 1.7 nm or 6 monolayers (MLs), while the top interface has a width of approximately 2 nm (7 MLs). These values are similar to those reported previously for a sample grown at 625°C.[11] Since the interface roughness visible along the $[2\bar{1}\bar{1}0]$ direction is relatively low (rms of 0.06 nm similar to that of GaN buffer), we expect equally smooth interfaces along the $[\bar{1}2\bar{1}0]$ direction that the STEM image integrates over due to the thickness of the specimen. Therefore, we conclude that the dominant contribution to the estimated interface widths is due to intermixing during MBE growth. Most important for narrow quantum wells, both interface widths decrease to approximately 1.5 nm when the substrate temperature is reduced to 550°C.

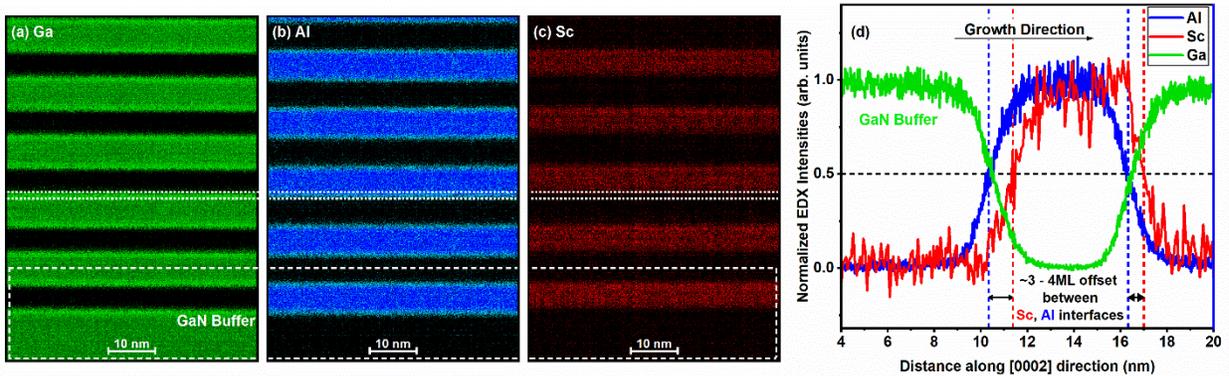

**Figure 5.** (a) Ga, (b) Al, and (c) Sc STEM EDX maps of a 6-nm-$Sc_{0.14}Al_{0.86}$N/6-nm-GaN SL grown at $T_{sub}$ = 625°C. The images correspond to the first five periods of the superlattice and have the GaN buffer visible at the bottom. (d) Elemental intensity profiles along the growth direction obtained by averaging over the width of the maps and normalized to a maximum of one, showing the onset and termination of the first $Sc_{0.14}Al_{0.86}N$ layer. The horizontal dashed line marks the half-point of the intensity while the vertical dashed lines indicate the position of the half-point intensity for both Ga and Al in blue, and for Sc in red.

This intermixing effect is further investigated with STEM EDX mapping of individual metal atoms. Figure 5 shows the STEM EDX maps of Ga (green), Al (blue), and Sc (red) of a 6-nm-GaN/6-nm-Sc$_{0.14}$Al$_{0.86}$N superlattice grown at $T_{sub}$ = 625°C, along with the corresponding elemental line profiles across the first Sc$_{0.14}$Al$_{0.86}$N layer calculated by averaging over the width of the images. The maps show overlapping interfaces, confirming interdiffusion across the interface. The half-points of Al and Ga intensities coincide and, since Ga dominates the HAADF-STEM signal, they also correspond to the points typically used to determine the width of the QWs and barriers (e.g. Fig. 4(b)). However, the line profiles indicate that Al reaches its maximum before Ga reaches its minimum, suggesting significant Ga intermixing in the barriers. Most surprisingly, the half-points of the Sc intensity at both onset and descent are delayed relative to the other metals by approximately 4 and 3 monolayers, respectively. The Sc-containing layer ends up about a monolayer thinner than, and shifted relative to the Al-containing layer. This suggests the two interfaces have different alloy distributions that include quaternary ScAlGaN. The delay in Sc incorporation may be due to displacement by Ga atoms intermixed into the barrier layer, as the Sc intensity reaches its maximum only after the Ga reaches its minimum. Ga may also prevent prompt incorporation of Sc atoms left on the surface after the shutter closed leading to the delay in the descent of the Sc signal.

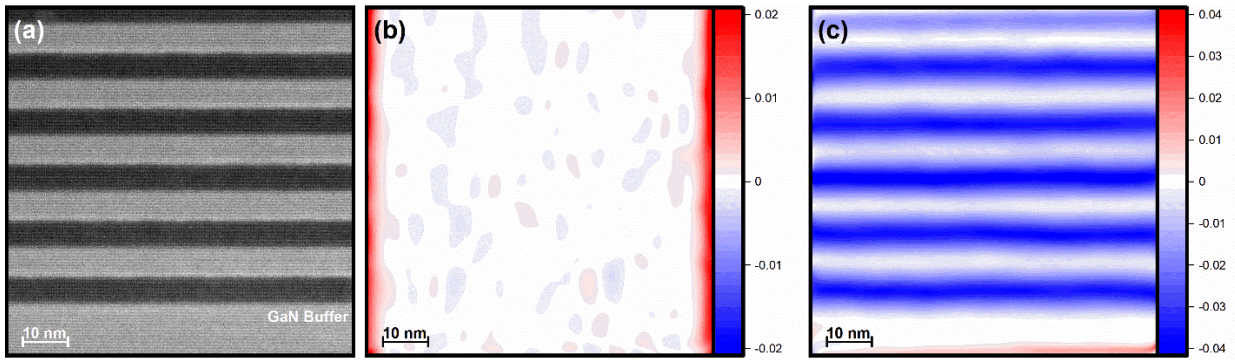

**Figure 6**. (a) HAADF STEM image of a 6-nm-Sc$_{0.14}$Al$_{0.86}$N/6-nm-GaN SL grown at 600°C substrate temperature, and its corresponding GPA maps for in-plane lattice strain (b), and out-of-plane lattice strain (c). The image corresponds to the first five periods of the superlattice and has the GaN buffer in view. The lattice strain is computed with respect to the buffer layer.

Geometric Phase Analysis (GPA) applied to HAADF-STEM images offers a powerful approach for quantitative strain mapping at the nanometer scale. GPA utilizes Fourier analysis of the periodic lattice fringes in STEM images to extract local structural information corresponding to selected reciprocal lattice vectors. By comparing the phase shifts relative to a reference region—typically chosen from a relaxed or well-characterized portion of the sample—local displacement and strain fields can be quantitatively determined with high spatial resolution. In this study, the in-plane ($\varepsilon_{xx}$) and out-of-plane ($\varepsilon_{yy}$) strain components of a 6-nm-$Sc_{0.14}Al_{0.86}N$/6-nm-GaN superlattice grown at a substrate temperature of 600°C were evaluated and are shown in Figure 6. The images capture the first five periods of the superlattice and include the underlying GaN buffer layer, which was used as the strain-free reference. Positive strain values (red) indicate larger inter-atomic spacing relative to GaN, whereas negative values (blue) indicate smaller spacing. The in-plane map reveals negligible mismatch between GaN and $Sc_{0.14}Al_{0.86}N$, confirming pseudomorphic growth. The same GPA analysis conducted near the 20th repeat of the superlattice showed similar negligible strain distribution (see Figure S7 in Supplementary Material). In contrast, GPA of a 6-nm-$Sc_{0.18}Al_{0.82}N$/6-nm-GaN SL exhibiting plastic relaxation in XRD RSM shows alternating in-plane strains that increase in amplitude with the number of SL periods (Figure S8). The out-of-plane strain component ($\varepsilon_{yy}$) displays periodic modulation, consistent with the alternating c-lattice parameters of the GaN and ScAlN layers. The $Sc_{0.14}Al_{0.86}N$ and $Sc_{0.18}Al_{0.82}N$ layers exhibit average out-of-plane lattice constant differences to GaN of approximately 4% and 3%, respectively, consistent with values previously observed in single-layer $Sc_xAl_{1-x}N$ films on GaN.[11]

While substrate temperatures around 600°C clearly improve overall structural uniformity and optical properties[25] of $Sc_{0.14}Al_{0.86}N$/GaN SLs with 6 nm GaN wells, they are not adequate for the growth of SLs with ultra-thin GaN layers. Due to enhanced intermixing at high temperature, GaN layers narrower than the interface width end up consisting of an inhomogeneous alloy rather than pure GaN.[11] Increased interface roughness has also been observed in $Sc_{0.14}Al_{0.86}N$/GaN SLs with 2 nm GaN layers grown at 625°C compared to those grown at 536°C substrate temperature.[25] For these structures, non-uniform interdiffusion can also introduce nanoscale strain at the interfaces in a nominally lattice-matched structure, contributing to the interfacial roughening caused by $Sc_{0.14}Al_{0.86}N$. Since interface roughness and widths decrease with growth temperatures, SLs with quantum wells narrower than 3 nm benefit from lowering the growth temperature to approximately 550°C.[25] Figure 7 shows the HAADF STEM structure and metal EDX maps for a 6-nm-$Sc_{0.14}Al_{0.86}N$/2-nm-GaN SL grown at 550°C. The elemental profiles through one QW and adjacent barriers calculated by averaging over the width of the images are plotted in Figure 7(e). At this growth temperature, the delay of the onset and termination of the Sc- relative to the Al-containing alloy is reduced to less than one monolayer. It is clear that intermixing still plays a major role in defining the alloy, and implicitly the band structure variation along the growth direction. Additional research needs to be done on growth procedures (e.g. shutter sequences, growth pauses) to further reduce interface widths to the monolayer level preferable for ultra-thin quantum wells.

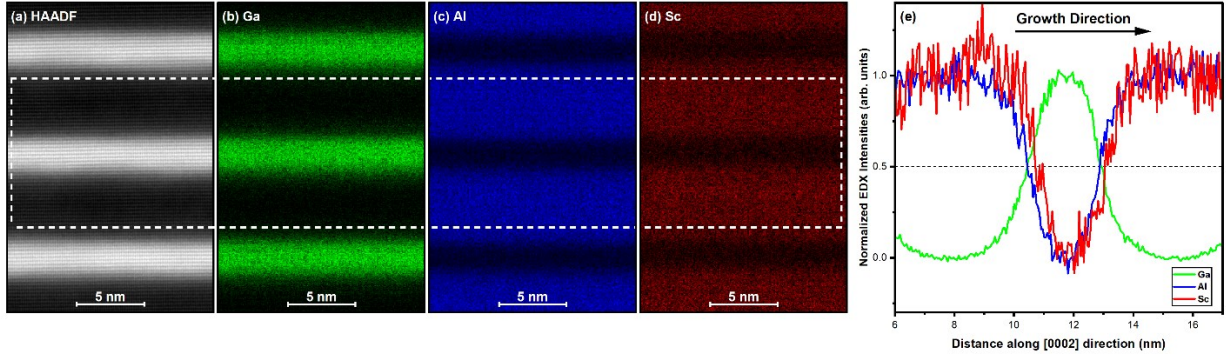

**Figure 7.** (a) HAADF STEM image of a 6-nm-$Sc_{0.14}Al_{0.86}N$/2-nm-GaN SL grown at 550°C and corresponding STEM EDX maps for (b) Ga, (c) Al, (d) Sc . (e) Elemental intensity profiles along the growth direction obtained by averaging over the width of the maps in the highlighted rectangle and normalized to a maximum of one. The horizontal dashed line marks the half-point of the EDX intensity normalized to a maximum of one.

The subtle dependence of the SL structure on growth temperature is due to the dissimilar MBE growth conditions for GaN and ScAlN. $Sc_xAl_{1-x}N$ layers are grown under metal-poor conditions that gradually increase the surface roughness during growth. Moreover, the roughness at the top of the $Sc_xAl_{1-x}N$ layer is found to increase with substrate temperature.[25] On the other hand, GaN is grown under metal-rich conditions and tends to smoothen the surface. A thick enough GaN layer on top of the rough $Sc_xAl_{1-x}N$ layer can periodically recover a smooth growth front and establish steady-state conditions that can be maintained throughout the SL growth. Conversely, thinner GaN layers cannot completely compensate for the roughening effect of ScAlN. As a result, in SLs with nanometer-thin GaN layers grown at elevated substrate temperatures, roughness accumulates with each SL period, leading to overall higher interface roughness.[11] Therefore, selecting the appropriate growth temperature is critical and depends on the relative thicknesses of $Sc_xAl_{1-x}N$ and GaN layers, as well as the intended application of the structure.

## IV. CONCLUSION

We have systematically studied the effect of Sc composition, GaN layer thickness, and substrate temperature on the structure of $Sc_xAl_{1-x}N$/GaN superlattices grown by molecular beam epitaxy. Our results confirm the lattice-matching condition for $Sc_xAl_{1-x}N$ on GaN occurs at 0.14±0.01 Sc composition. $Sc_{0.14}Al_{0.86}N$/GaN SLs are coherent with the underlying GaN substrate regardless of the thickness of GaN layers and show superior structural and optical properties. However, if desired, the relaxation critical thickness of compressively strained ScAlN/GaN SLs can be increased considerably beyond the single ScAlN film limit with thick GaN interlayers. In-depth structural examination with STEM identified interface width as a critical parameter to consider when tailoring growth conditions for optimal functionality. EDX also evidenced a subtle interaction between the three metal components at interfaces. GPA provided additional validation of the lattice-matching condition and valuable complementary information about nanoscale strain distribution in lattice-mismatched superlattices. Substrate temperature of 600-625°C is found to be optimal for the growth of $Sc_{0.14}Al_{0.86}N$/GaN SLs with layer thicknesses of 6 nm for each constituent layer. However, the growth temperature needs to be lowered for SLs with extremely thin (≤2 nm) GaN layers if the device application relies on precise modulation of the alloy composition and band edge profile. These findings provide valuable insights for the design and growth of lattice-matched $Sc_{0.14}Al_{0.86}N$/GaN multilayers with excellent structural quality required for fabrication of novel photonic devices.

## SUPPLEMENTARY MATERIAL

Supplementary Material contains additional XRD, AFM, and HAADF-STEM data.

## ACKNOWLEDGEMENTS

We acknowledge support from the National Science Foundation (NSF). G. G., Z. U. A., and O.M. acknowledge partial support from NSF award DMR-2414283. All STEM imaging and analyses were performed at the Electron Microscopy Facility at the Birck Nanotechnology Center, Purdue University.

## CONFLICT OF INTEREST STATEMENT

The authors have no conflicts to disclose.

## DATA AVAILABILITY

The data that supports the findings of this study are available within the article.

# Supplementary Material

# Structural optimization of lattice-matched Sc$_{0.14}$Al$_{0.86}$N/GaN superlattices for photonic applications

Rajendra Kumar[1,2], Govardan Gopakumar[1,2], Zain Ul Abdin[1,2], Michael J. Manfra[1,2,3,4], and Oana Malis[1,2*]

[1]Dept. of Physics and Astronomy, Purdue University, West Lafayette, IN USA 47907

[2]Birck Nanotechnology Center, West Lafayette, IN USA 47907

[3]School of Materials Engineering, Purdue University, West Lafayette, IN USA 47907

[4]Elmore Family School of Electrical and Computer Engineering, Purdue University, West Lafayette, IN USA 47907

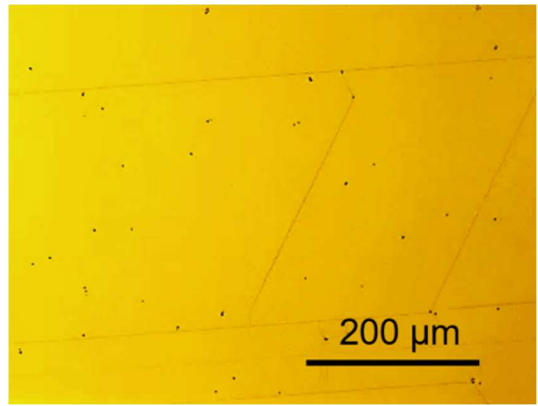

**Figure S1.** Optical micrograph of 50-period 6-nm-Sc$_{0.12}$Al$_{0.88}$N/6-nm-GaN SL exhibiting cracks due to relaxation of tensile strain generated by lattice mismatch between SL and GaN.

---

[*] Author to whom correspondence should be addressed. Electronic mail: omalis@purdue.edu

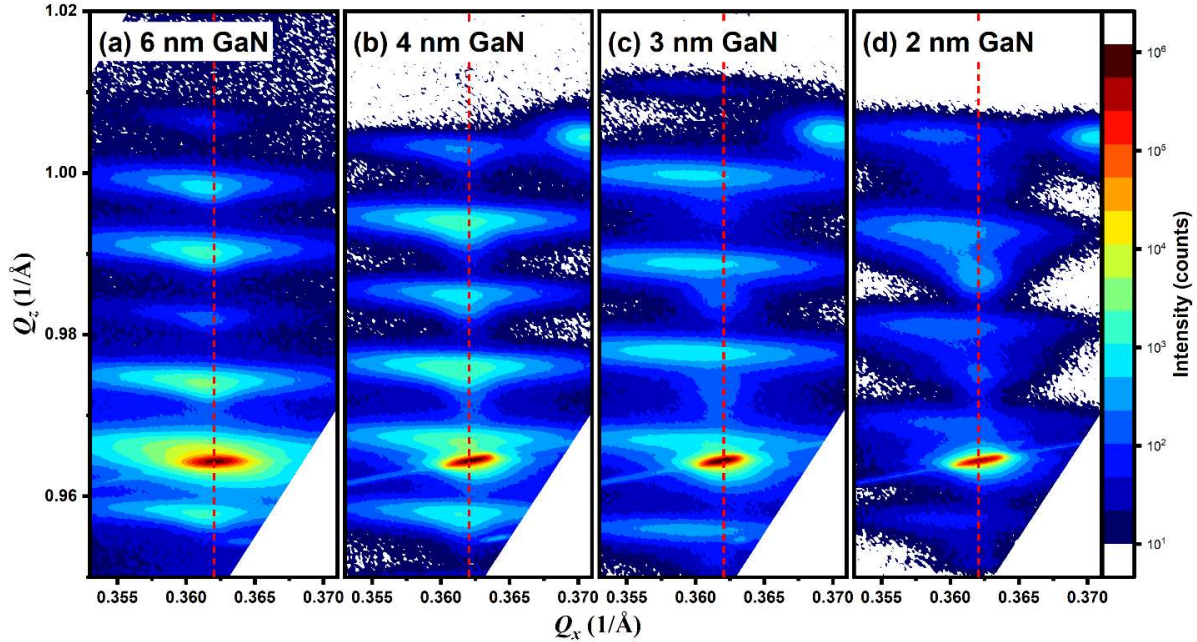

**Figure S2.** XRD RSMs of the $(10\bar{1}5)$ reflection for 50-period 6-nm-$Sc_{0.18}Al_{0.82}N$/GaN SLs with GaN layer thicknesses of (a) 6 nm, (b) 4 nm, (c) 3 nm, and (d) 2 nm. The position of the SL peaks shifts to lower $Q_x$ values with decreasing GaN thickness, indicating an increased degree of relaxation. The growth temperatures for the sample in panels (a)-(d) are 580°C, 515°C, 515°C, and 525°C, respectively.

**Table S1.** Average in-plane lattice mismatch between SL and GaN substrate calculated from the XRD RSM maps in Figure S2 for $Sc_{0.18}Al_{0.82}N$/GaN SL samples with various GaN thicknesses. For the SL with 2 nm GaN thickness, the satellites corresponding to the relaxed region are used to calculate lattice mismatch.

| GaN layer thickness | *a* lattice mismatch between SL and GaN substrate |
|---|---|
| 6 nm | 0.26 % |
| 4 nm | 0.34 % |
| 3 nm | 0.66 % |
| 2 nm | 1.02 % |

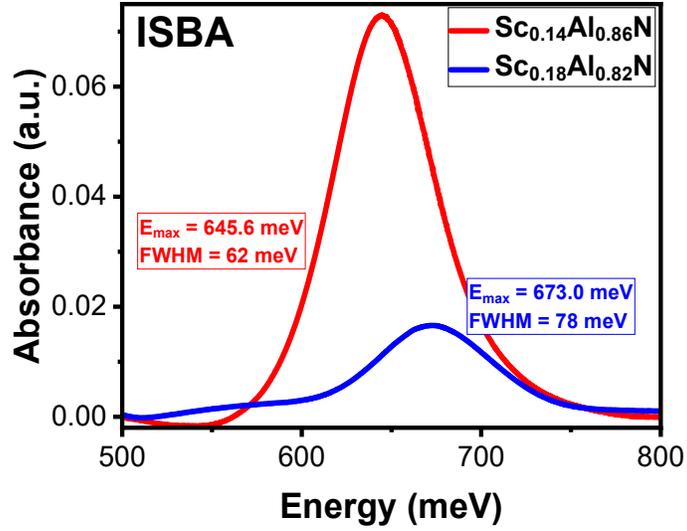

**Figure S3.** Infrared intersubband absorption comparison between a lattice-matched $Sc_{0.14}Al_{0.86}N$/GaN SL and a lattice-mismatched $Sc_{0.18}Al_{0.82}N$/GaN SL with 2 nm GaN layers and 6 nm ScAlN layers.

**Table S2**: FWHM of the line profile along $Q_x$ and $Q_z$ directions in XRD RSM for three SL peaks of the 6-nm-$Sc_{0.14}Al_{0.86}N$/6-nm-GaN SLs grown at different temperatures shown in Figure 3.

| $T_{sub}$ | Vertical Profile (along $Q_z$, µm⁻¹) | | | Horizontal Profile (along $Q_x$, µm⁻¹) | | |
|---|---|---|---|---|---|---|
| | Peak 1 ($Q_z \sim 1.01$) | Peak 2 ($Q_z \sim 1.00$) | Peak 3 ($Q_z \sim 0.99$) | Peak 1 ($Q_z \sim 1.01$) | Peak 2 ($Q_z \sim 1.00$) | Peak 3 ($Q_z \sim 0.99$) |
| 510°C | 14.3 | 15.2 | 15.3 | 18.8 | 18.3 | 18.4 |
| 550°C | 8.7 | 9.0 | 8.9 | 13.7 | 13.6 | 13.6 |
| 600°C | 7.6 | 7.1 | 6.6 | 11.7 | 11.1 | 11.3 |
| 625°C | 6.8 | 6.5 | 6.1 | 11.2 | 11.1 | 11.1 |

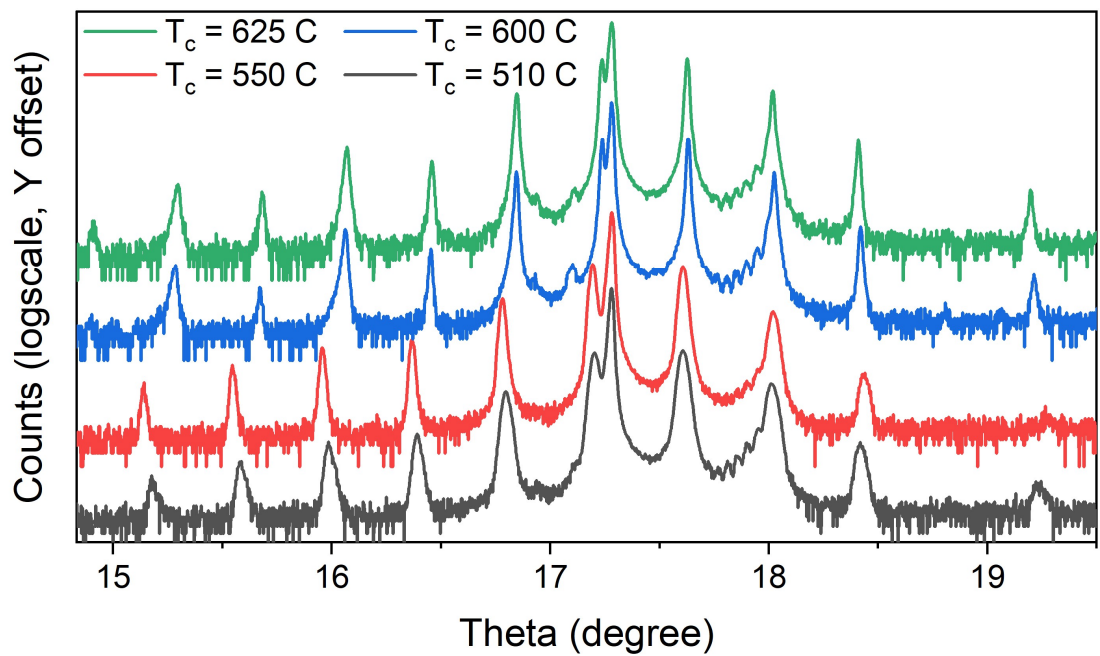

**Figure S4.** HRXRD ω-2θ scans of the 6-nm-$Sc_{0.14}Al_{0.86}N$/6-nm-GaN superlattices grown at various substrate temperatures corresponding to Figure 3.

**Table S3.** FWHM of the most intense SL peak and the GaN substrate peak from fitting the HRXRD ω-2θ scans of the $Sc_{0.14}Al_{0.86}N$/GaN SL samples shown in Figure S4.

| Substrate temperature (°C) | FWHM (arcseconds) | |
|---|---|---|
| | SL peak | GaN substrate |
| 510 | 142 | 57 |
| 550 | 109 | 59 |
| 600 | 49 | 48 |
| 625 | 56 | 61 |

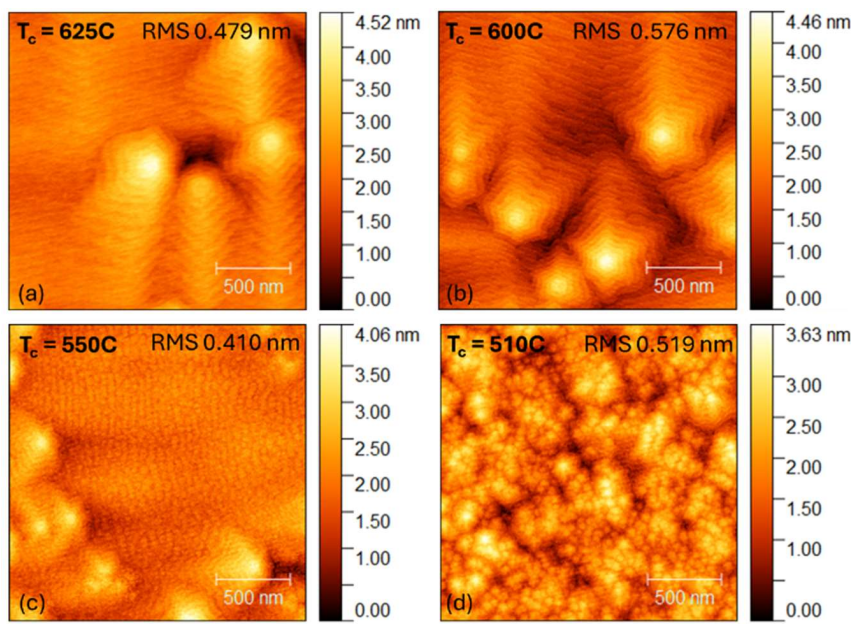

**Figure S5.** AFM images of 6-nm-Sc$_{0.14}$Al$_{0.86}$N/6-nm-GaN SLs grown at different substrate temperatures. SL samples grown at 600°C and 625°C exhibit clear atomic step morphology.

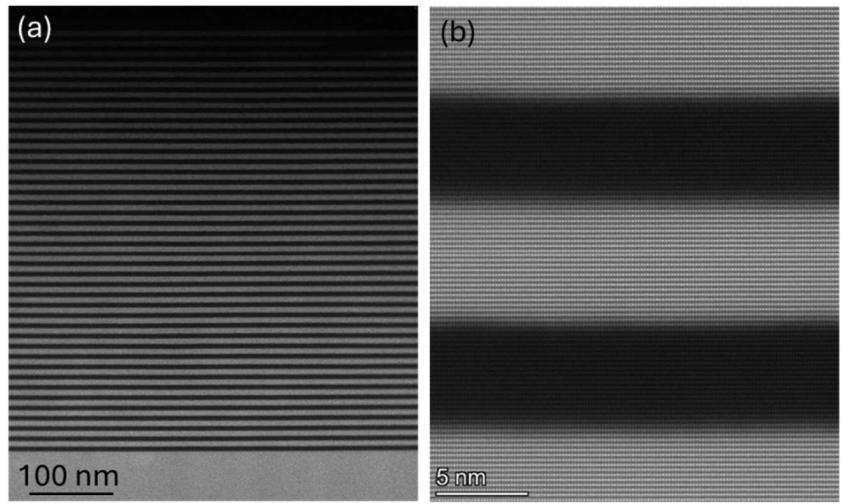

**Figure S6**. Low magnification (a) and high-magnification (b) HAADF-STEM images of 6-nm-Sc$_{0.14}$Al$_{0.86}$N/6-nm-GaN superlattice grown at $T_{sub}$ = 625°C exhibiting perfectly lattice-matched structure.

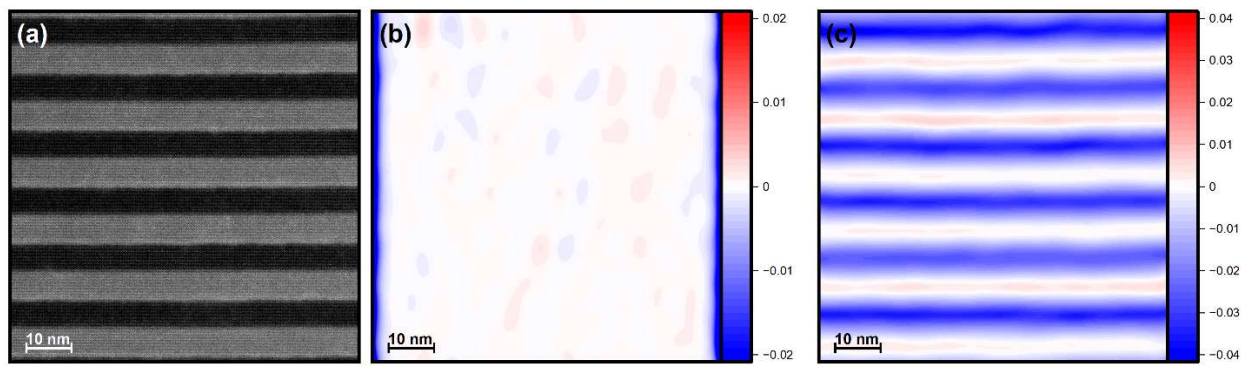

**Figure S7.** (a) HAADF-STEM image, (b) in-plane lattice mismatch map, and (c) out-of-plane lattice mismatch map of a 6-nm-GaN/6-nm-Sc$_{0.14}$Al$_{0.86}$N superlattice, grown at $T_{sub}$ = 600 °C. The image corresponds to the middle of the superlattice (near the 20$^{th}$ period). The lattice mismatch is computed with respect to the GaN layers in the superlattice.

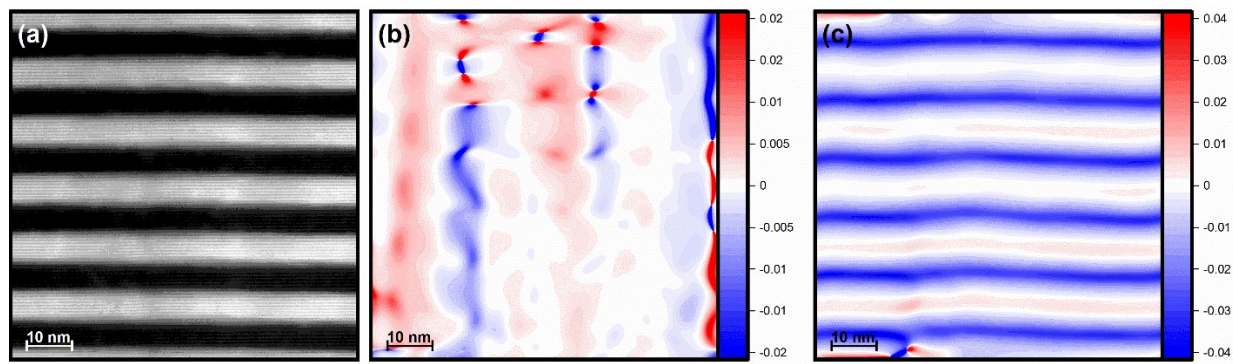

**Figure S8.** (a) HAADF-STEM image, (b) in-plane lattice strain map, and (c) out-of-plane lattice strain map of a 6-nm-Sc$_{0.18}$Al$_{0.82}$N/6-nm-GaN superlattice grown at $T_{sub}$ = 550°C. The image corresponds to the middle of the superlattice. The lattice strain is computed relative to the lattice constants of the GaN layers in the superlattice.

Figure S8 shows the HAADF-STEM image and lattice strain maps for a 6-nm-Sc$_{0.18}$Al$_{0.82}$N/6-nm-GaN superlattice grown at $T_{sub}$ = 550°C. This sample exhibits pronounced in-plane lattice mismatch, reflecting strain accumulation. Our prior study[12] demonstrated that Sc$_{0.18}$Al$_{0.82}$N films grown on GaN undergo plastic deformation to relax to larger in-plane lattice constants when their thickness exceeds the critical limit for pseudomorphic growth. As discussed earlier, alternating GaN layers in the superlattice can partially mitigate plastic relaxation, but nanoscale deviations of the in-plane lattice constant that are consistent with strain fields originating from misfit dislocations are evident in the in-plane strain map. Blue regions in the in-plane map correspond to the insertion of extra atomic planes, whereas red regions indicate plane terminations. The distortions correlate in the growth (vertical) direction and alternate from compressive to tensile in-plane (horizontal direction) with a periodicity of approximately 50 nm. The maximum distortions also increase from bottom to top of the image as expected from strain accumulation with the number of SL periods.